# Negative Differential Resistance in Boron Nitride Graphene Heterostructures: Physical Mechanisms and Size Scaling Analysis

Y. Zhao[1], Z. Wan[1], X. Xu[1], S. R. Patil[1, 2], U. Hetmaniuk[3], and M. P. Anantram[1]

[1]Department of Electrical Engineering, University of Washington, Box 352500, Seattle, WA 98105-2500, USA
[2]Department of Physics, College of Engineering, Pune, Maharashtra, 411005, India
[3]Department of Applied Mathematics, University of Washington, Box 353925, Seattle, WA 98105-3925, USA

Hexagonal boron nitride (hBN) is drawing increasing attention as an insulator and substrate material to develop next generation graphene-based electronic devices. In this paper, we investigate the quantum transport in heterostructures consisting of a few atomic layers thick hBN film sandwiched between graphene nanoribbon electrodes. We show a gate-controllable vertical transistor exhibiting strong negative differential resistance (NDR) effect with multiple resonant peaks, which stay pronounced for various device dimensions. We find two distinct mechanisms that are responsible for NDR, depending on the gate and applied biases, in the same device. The origin of first mechanism is a Fabry-Pérot like interference and that of the second mechanism is an in-plane wave vector matching when the Dirac points of the electrodes align. The hBN layers can induce an asymmetry in the current-voltage characteristics which can be further modulated by an applied bias. We find that the electron-phonon scattering introduces the decoherence and therefore suppresses first mechanism whereas second mechanism remains relatively unaffected. We also show that the NDR features are tunable by varying device dimensions. The NDR feature with multiple resonant peaks, combined with the ultrafast tunneling speed provides prospect for the graphene-hBN-graphene heterostructure in the high-performance electronics.

## 1. Introduction

Graphene, a two-dimensional material with unique mechanical, thermal and electronic transport properties [1] is a promising candidate for nanodevices as it is deeply scaled in one dimension and the lithography offers scaling in the other two dimensions. Building devices based on graphene is, however, partially impeded by the lack of compatible insulating substrate. Hexagonal boron nitride (hBN) has an atomically smooth two dimensional (2D) layered structure with a lattice constant very similar to that of graphene (1.8% mismatch), sufficiently large electrical band gap (~4.7eV), and excellent thermal and chemical stability [2], allowing it to be stacked with graphene to build device structures with desired functionalities. Also, hBN reduces the surface roughness of graphene without degrading its giant mobility. [3,4] The nano-scale devices based on graphene employing atomically thin hBN with novel electrical and optical properties have recently been reported. [5-15] An appearance of negative differential resistance (NDR) in such devices further interests the researchers as it could potentially impact the number of applications such as high-speed IC circuits, signal generators, data storage, and so on. [16]

NDR in double barrier resonant tunneling diodes (DB-RTD), appears when the quasi-bound levels can no longer enhance the tunneling resonantly. [17] Recent theoretical investigations report the appearance of NDR features in pure graphene based devices, involving nanoribbon superlattice [18], doped junctions [19-23], tunnel-FET [24,25], and MOSFET structures [26]. These structures typically employ graphene with fine-tuned bandgap, such that graphene behaves more like a semiconductor. NDR effect is also being reported in a single as well as multilayer heterostructure of graphene-hBN-graphene. [27-29] Reference [30] models NDR peak in a near metallic bi-layer graphene device. Apart from this, such devices could also find applications in multi-valued memory. [31,32] The multilayer graphene-hBN-graphene heterostructure based electronic devices particularly attract the attention of engineers due to their relatively simpler fabrication. [33-35]

The NDR features in multilayer based devices are to be investigated by exploring current-voltage characteristics as a function of (i) number of hBN layers, (ii) lateral dimensions in determining both the voltage location of NDR peaks and the peak-to-valley ratio, which are essential in the device design, (iii) the role of the asymmetric band offset between hBN and graphene, and (iv) defects and scattering. This, however, necessitates further research to rationalize the underlying physics of the NDR effect and gain insight on how to control its critical properties mentioned above.

In this work, we focus on a prototypical multilayer device is shown in Fig. 1(a), which consists of layers of graphene and hBN that are vertically stacked. The graphene layers serve as conducting electrodes with a unique band structure while the hBN layers are tunnel barriers. We model the electron transport in these devices by atomistic non-equilibrium Green's function (NEGF) method. A decoherence mechanism, the electron-phonon scattering is introduced and its impact on both NDR effects is presented. Additionally, we demonstrate how the magnitude of current, locations of resonant peaks, and peak-to-valley ratio (PVR) values can be tuned by the device parameters. The modeled devices range from a small system with 6,000 atoms to experimentally feasible sizes up to 70,000 atoms (lateral dimensions 24.6nm × 27nm).

Next, section 2 defines our method by discussing the underlying Hamiltonians and the methodology for the computation of the quantum transport. Section 3 demonstrates the results and discusses the NDR effects with two underlying mechanisms along with the role of electron-phonon scattering. Section 4 presents the size scaling analysis followed by section 5 concluding the work.

## 2. Method

A prototypical heterostructure consists of two semi-infinitely long monolayer armchair-edged graphene nanoribbon (AGNR) electrodes sandwiching an ultra-thin hBN film, with a vertically applied external gate electric field as shown in Fig 1(a). AGNR is employed because it can be engineered as an intrinsic conductor. This forms a vertical tunneling heterostructure with hBN acting as a potential barrier. The hBN film is sandwiched between a bottom and top AGNR, forming a central overlapping *heterostructure/multilayer* region stacked in AB order (Bernal stacking). The lattice constant mismatch between hBN and graphene is negligibly small, 1.8%, therefore, we build the device structure with the uniform lattice constant of graphene (2.46 Å) only. The system Hamiltonian is constructed using the nearest neighbor tight binding approximation, with the parameters [11,36]: $E_{on-site}^{C} = 0$, $E_{on-site}^{B} = 3.34\text{eV}$, $E_{on-site}^{N} = -1.4\text{eV}$, and $t_{intralayer}^{C-C} = 2.64\text{eV}$, $t_{intralayer}^{B-N} = 2.79\text{eV}$, $t_{interlayer}^{B-N} = 0.60\text{eV}$, $t_{interlayer}^{C-B/N} = 0.43\text{eV}$. Only the low energy $p_z$ orbitals are considered here; so that the Hamiltonian has the same dimension as the total number of atoms simulated. The effect of number of tunneling hBN layers ($N_z$), the system width ($N_x$) and the length of the multilayer stacking region ($N_y$), where units of $N_x$ and $N_y$ are number of atoms, on the device performance is investigated. The nanostructure thickness is ($N_z + 2$) in units of atomic layers, which includes the two monolayer graphene sheets at the ends.

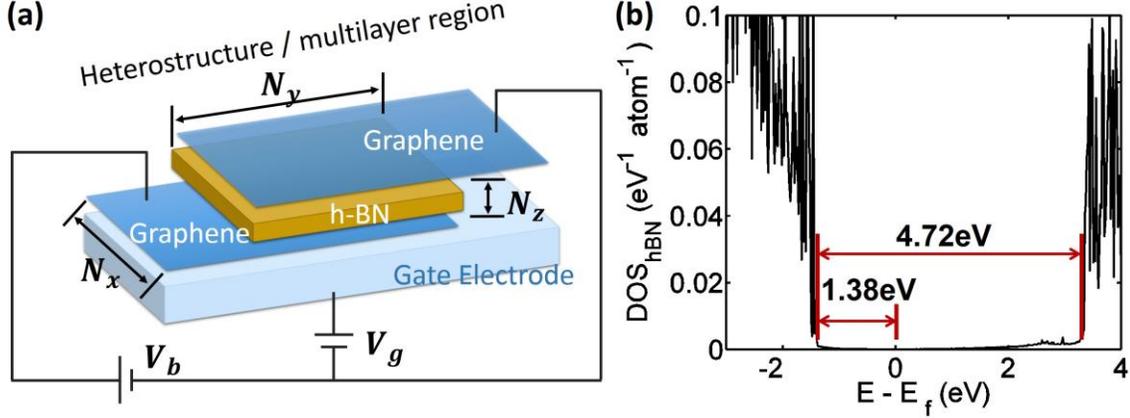

Figure 1. (a) A schematic view of the heterostructure device. $N_x$ and $N_y$ represent the width and stacking length of the device respectively. $N_z$ is the number of hBN layers sandwiched between the two AGNR ribbons. All the dimensions are in unit of atomic layers. (b) The average DOS versus Energy of hBN for device with $N_x = 200$, $N_y = 32$ and $N_z = 3$. This shows a 4.72eV bandgap of atomically thin hBN material and a 1.38eV valence band-offset between graphene and hBN stacking structure.

The bias across the heterostructure is applied by rigidly shifting the electrostatic potential of the bottom graphene electrode by the amount equal to the applied bias as the metallic graphene layers have much higher conductivity than hBN. The electrostatic potential energy of the bottom layer is $U = -eV_b$, where $V_b$ is the applied bias, and the electrostatic potential of the top graphene layer remains zero. The electrostatic potential at each of the sandwiched hBN layers are determined by linearly increasing/decreasing the potential from top to the bottom graphene layer, because the $c$-axis (out of plane) conductivity of hBN is orders of magnitude smaller than the in-plane conductivity of graphene. The chemical potential of contacts are controlled by bias voltage, namely $\mu_B = -eV_b$ and $\mu_T = 0$ for bottom and top graphene leads respectively. The gate voltage is modeled by shifting the electrostatic potential at the bottom electrode by $\Delta U = -0.01eV_g$, where $V_g$ is the gate voltage. We choose $N_x$ equal to $3n + 2$, where $n$ is an arbitrary positive integer, [37,38] such that the AGNR have zero bandgap. All calculations were performed at 300K.

We simulate the transport properties of the device by using the state of the art Green's function (NEGF) formalism. [39] After the generation of real-space Hamiltonian matrix, the retarded Green's function is computed by

$$\mathbf{G}^r(E) = [E\mathbf{I} - \mathbf{H} - \mathbf{\Sigma}_L^r(E) - \mathbf{\Sigma}_R^r(E)]^{-1}$$

where $E$ is the electrons energy, $\mathbf{I}$ denotes the identity matrix, $\mathbf{H}$ is the system Hamiltonian matrix, and $\mathbf{\Sigma}_{L/R}^r$ is the self-energy capturing the semi-infinite left (right) graphene sheets shown in Fig. 1. The local density of states at a given atom site $i$ can be calculated subsequently by

$$\text{DOS}(E, i) = -\frac{1}{\pi}\text{Im}\{\text{trace}[\mathbf{G^r}(E)_{i,i}]\}$$

The transmission coefficient is then calculated using the expression,

$$T(E) = \text{trace}[\mathbf{\Gamma_L}(E)\mathbf{G^r}(E)\mathbf{\Gamma_R}(E)\mathbf{G^{r\dagger}}(E)]$$

where $\mathbf{\Gamma_L}(E) = i[\mathbf{\Sigma_L^r}(E) - \mathbf{\Sigma_L^{r\dagger}}(E)]$ and $\mathbf{\Gamma_R}(E) = i[\mathbf{\Sigma_R^r}(E) - \mathbf{\Sigma_R^{r\dagger}}(E)]$, and the current as a function of bias is given by,

$$I = \frac{2e}{\hbar}\int \frac{dE}{2\pi} T(E)[f(E - \mu_R) - f(E - \mu_L)]$$

where $f(E)$ is the Fermi function and $\mu_{L/R}$ represent the Fermi level of the left (top) and right (bottom) graphene monolayer electrode. In Fig. 1(b), we plot the DOS for a structure with a width of $N_x = 200$, $N_y = 32$, and with three hBN layers $N_z = 3$, at zero bias. The bandgap of hBN is found to be around 4.72eV, and the valence band offset between hBN and graphene is around 1.38eV, which is consistent with the prior work [40].

Traditionally, the recursive Green's function (RGF) approach is applied to solve for the retarded Green's function. Here, an algorithm named HSC-extension is developed, which is more efficient and is based on the nested dissection method, so that the requisite large scale systems can be simulated. [41]

## 3. Mechanisms

### 3A. Origin of Multiple NDR peaks (Mechanism 1)

Fig. 2(a) presents the computed current-voltage characteristics of the heterostructure with a transverse width of $N_x = 62$ (7.6nm), stacking length $N_y = 32$ (6.8nm) and three hBN layers (1.4nm) serving as the tunneling barrier. First we consider the highlighted curve with $V_g = 0$, in which case two NDR peaks emerge at $V_b = 0.3\text{V}$ and 0.66V, respectively. We attribute the formation of these multiple NDR peaks to the Fabry-Pérot like interference in the multilayer region (*mechanism 1*), as rationalized below.

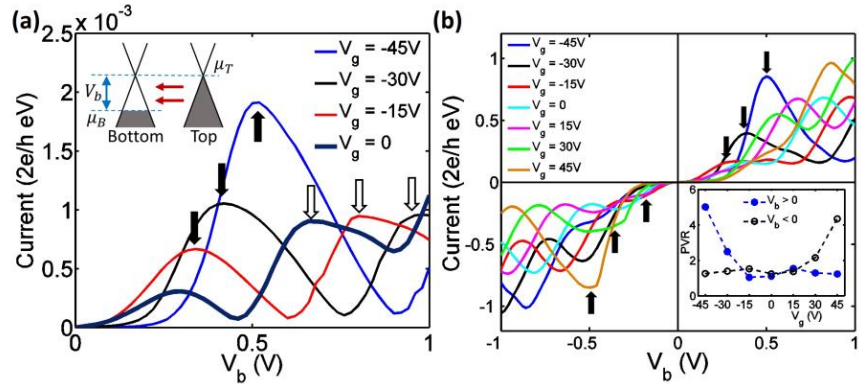

Figure 2. (a) Current versus drain voltage for a device with $N_x = 62$, $N_y = 32$ and $N_z = 3$. $V_g$ varies from -45V to 0. The black solid arrows in the four plots mark the current resonant peaks due to mechanism 2, and the empty arrows marks the NDR peak due to mechanism 1. The inset explains the resonant tunneling induced from mechanism 2. The difference

between Fermi energy and Dirac point in bottom graphene is induced by gate potential. When $V_b = -0.01V_g$, the electronic spectra of top and bottom electrodes are tuned into alignment, allowing the resonant tunneling. (b) Current versus drain voltage for large device with $N_x = 200$, $N_y = 32$ and $N_z = 1$. Here $V_g$ varies from -45V to +45V. Inset shows an asymmetric PVR relationship with the applied vertical gate potential.

To understand the multiple NDR peaks, we calculate the transmission and the average density of states ($DOS_g$ and $DOS_{hBN}$) at $V_b = 0.3V$, 0.46V and 0.66V, corresponding to the first peak, the first valley and the second peak in the current-voltage characteristics. In the $DOS_g$ curves (Fig. 3), the average DOS of bottom and top graphene layers are plotted separately. In particular, the huge peaks in $DOS_g$ are marked as peak **S**, which captures the edge states due to the zigzag-shaped cut-ends of graphene ribbons. The blue $DOS_{hBN}$ curves show the average DOS of the three hBN barrier layers. For the sake of comparison, the transmission coefficient at equilibrium is also plotted with the black dashed curves. The chemical potentials of the bottom and top AGNR are marked as vertical black lines ($\mu_B$ and $\mu_T$). The transmission and $DOS_g$ show a strong Fabry-Pérot like resonant feature in the low energy window. The semi-infinite top and bottom AGNRs couple with hBN at the central heterostructure (multilayer) region. The potential discontinuity caused by the interaction between the hBN cut-ends and the graphene layers create a resonant cavity in the overlapping region at both the top and bottom graphene layer. When electrons transport across the boundaries between graphene monolayer and hBN multilayer regions, partial reflections occur at the interfaces. As a result, the transmission is oscillatory with peaks and valleys corresponding to constructive and destructive interferences.

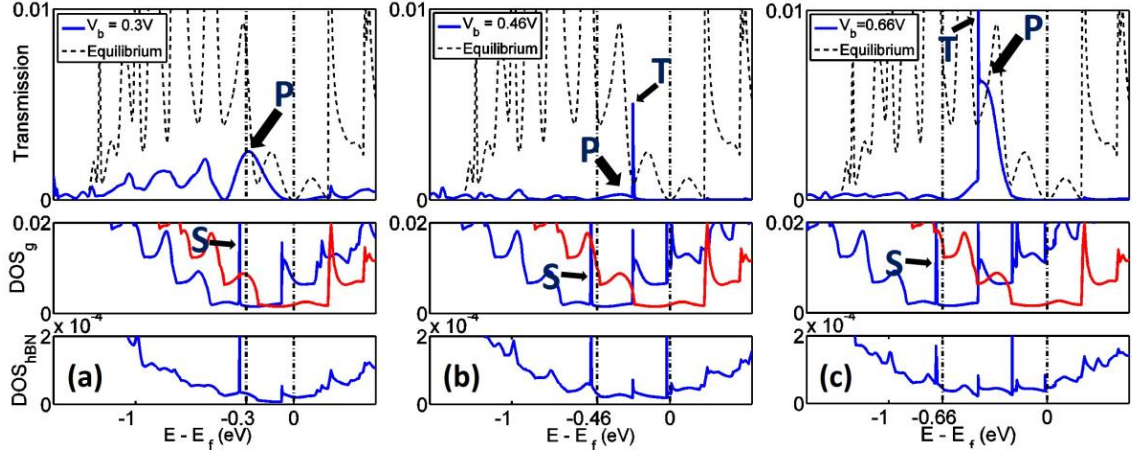

Figure 3. Transmission and DOS plot at various biases for I-V curve ($V_g = 0$) in Fig. 2(a). (a)-(c) specifies the bias potential $V_b = 0.3V$, 0.46V and 0.66V respectively. In the transmission plots, black dashed curves are transmission coefficient when $V_b = 0$. In $DOS_g$ plots, blue and red curves represent DOS of bottom and top graphene sheets respectively. Vertical dash-dot lines give the chemical potentials at both graphene ends $\mu_B$ and $\mu_T$, which determines the bias window. **S** mark the DOS peaks resulting from the zigzag shaped edges of graphene cut ends. **P** mark the transmission peaks that mainly contribute to the current. **T** represent the tunneling peaks due to the energy alignment of subbands in top and bottom graphene contacts; they do not contribute significantly to current. Units of DOS are number of states per atom per eV.

The current is determined from the area enveloped by the transmission curve in the energy window bounded by the Fermi levels of two electrodes, $\mu_B$ and $\mu_T$ (black dash-dot lines in Fig. 3). The transmission peak (**P**) at $E = -0.3$eV in Fig. 3(a) mainly contributes to the tunneling current. At low bias regime ($V_b < 0.18V$), we find that this transmission peak **P** is enhanced, resulting in the increase of current with applied bias. The resonant tunneling occurs when the constructive quantum interference assists the tunneling of electrons from the top to bottom electrodes at specific energies. When the bias is further increased (until 0.46V), the transmission peak **P** is reduced due to destructive interference despite the fact that the energy window for carrying current enlarges. This transmission reduction begins to dominate after $V_b = 0.3$V, which induces a drop in current. At $V_b = 0.46$V, there is a large suppression of transmission within the bias window, which creates a large tunneling gap, leading to the current valley as reflected in the highlighted curve of Fig. 2(a). Note that the density of states is large in both graphene electrodes even when the transmission is small as see in Fig. 3(b). Then, the transmission starts to increase again at around $V_b = 0.66$V due to the constructive interference.

An interesting feature of Fig. 3(a) is that only one transmission peak is observed at around $E = -0.3$eV ($\mu_T$), while a symmetric peak at $E = 0$eV ($\mu_B$) is clearly absent. This is due to the fact that the presence of hBN layers break the symmetry. We could understand this from DOS$_{hBN}$ plot at $V_b = 0.3$V (Fig. 3(a) DOS$_{hBN}$ curve), which shows a peak near $E = -0.3$eV but a valley at $E = 0$eV. This means that electrons at $E = 0$eV see a stronger barrier when tunneling between AGNR layers, and suggests the breaking of the $\pi - \pi^*$ symmetry. This argument is tested by considering a symmetric tunnel barrier, where such an asymmetry in transmission does not exist. In Fig. 3(b) and (c), sharp transmission peaks (marked as peak **T**) are observed. This significant tunneling enhancement results from the energy level alignment between the subbands of top and bottom graphene electrodes, as reflected in the corresponding DOS$_g$ features.

### *3B. Gate induced NDR peak (Mechanism 2)*

We next discuss the second mechanism that leads to single intense NDR peak by investigating the operational behavior of the heterostructure in the presence of an external gate voltage ($V_g$). Fig. 2 shows the current-voltage characteristics for a family of $V_g$ ranging from -45V to +45V for two devices with different sizes. Take the current-voltage curve at $V_g = -45$V as an example; At $V_b = 0$, the negative gate voltage shifts the energy of Dirac point in the bottom AGNR electrode to $U = 0.45$eV at equilibrium, while preserving the chemical potentials from two contacts at $\mu_B = \mu_T = 0$. At $V_b = 0.45$V (see Fig. 2(a) inset), the Dirac points of bottom and top AGNR electrodes are aligned. As a result, electrons can tunnel from the valence band of the top graphene layer to the conduction band of the bottom graphene layer owing to the in-plane wave vector conservation [35]. This particular mechanism (*mechanism 2*) induces the resonant transmission and results in the large current peaks marked by solid arrows in Fig. 2. It is noticeable that current peak positions are shifted from the theoretical prediction ($V_b = -0.01V_g$) based on mechanism 2 only. This occurs when the strength of mechanism 2 is comparable to that of mechanism 1, when the voltage at which the peak current occurs is influenced by the Fabry-Pérot like interference. The superposition of mechanisms 1 and 2 leads to the current peak displacement, which is larger at low gate voltage (for example, $V_g = -15V$).

*3C. Gate response*

Gate voltage has different impacts on NDR induced by two distinct mechanisms discussed above. In Fig. 2(a), the peak current (solid arrows) increases with $V_g$ as the peak current is proportional to the number of carriers between $\mu_B$ and $\mu_T$ when Dirac points of the top and bottom graphene align (see inset of Fig. 2(a)). In contrast to this, we find that in Fig. 2(a), the peak current of the NDR peaks induced by mechanism 1 (empty arrows) are relatively insensitive to $V_g$. In addition, the PVR values of these NDR peaks are also $V_g$-insensitive because the vertical gate potential tunes the resonant energies for constructive interference, but do not affect the number of tunneling carriers. Consequently, for two types of NDR effects in a single device, the amplitudes of current peaks for mechanism 2 is weaker than that for mechanism 1 at low $V_g$, but can become significantly stronger at large gate voltage, as shown in Fig. 2(a) when $V_g = -45V$. When the device structure is enlarged to $N_x = 200$ and $N_y = 32$, the current-voltage curves for various gate voltages (Fig. 2(b)) show that the *multiple NDR peaks* stay clearly defined and their locations are strongly gate-controlled. We point out that the current-voltage curves are asymmetric for positive and negative biases even at $V_g = 0$ as the hBN layers breaks the $\pi - \pi^*$ symmetry in the multilayer system. We also note that after the NDR peak, our calculations clearly show a trend of increase in current with increase in drain voltage, in a manner qualitatively similar to the experiments. [35]

3D. Effect of scattering

It is to be noted that the mechanism 2 induced single current peak only occurs at specific bias points, determined by $V_g$. Besides, the amplitude of this single peak is $V_g$-sensitive, which helps us to distinguish it from the multiple peaks induced by mechanism 1. However, in the experiment [35] only strong resonant peaks due to mechanism 2 are observed. The disappearance of peaks originated from mechanism 1 merits the discussion.

In experiments on large area devices such as [35], unavoidable decoherence mechanisms such as electron-phonon, electron-electron and electron-environment interactions are non-negligible. To model decoherence, here we next introduce electron-phonon scattering in top and bottom graphene electrodes and investigate its influence on NDR peaks. The detailed modeling methodology can be found elsewhere [42]. The typical values of electron-phonon coupling constants and phonon energy are taken from [43]: elastic deformation potential $D_{el} = 0.01 eV^2$, inelastic deformation potential $D_{inel} = 0.07 eV^2$ and phonon energy $\hbar\omega = 180 meV$. The deformation potential values phenomenologically represent the strength of electron-phonon coupling, in terms of the electron mean free path which can be experimentally measured. Physically, larger $D_{el}$ and $D_{inel}$ represent stronger electron-phonon interaction and thus shorter electron mean free path. The simulated electron mean free path corresponding to the above parameters is about $1.48 \mu m$, which represents moderate scattering. The current-voltage curves are plotted in Fig. 4.

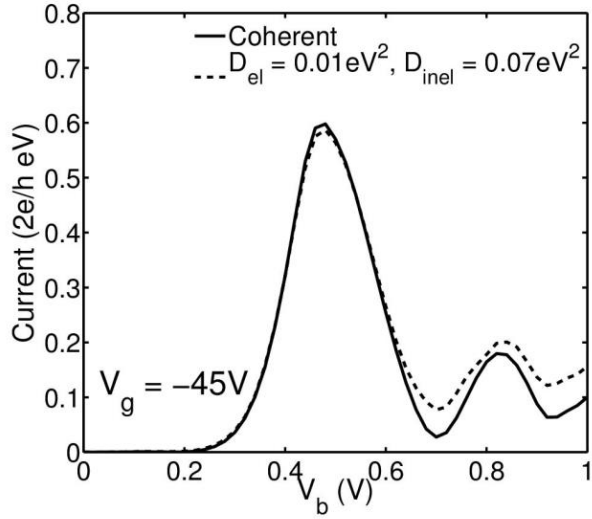

Figure 4. Current-voltage curves at $V_g = -45V$ for devices with $N_z = 1$, $N_x = 62$ and $N_y = 64$ with consideration of electron-phonon scattering. The deformational potentials used in the calculation are $D_{el} = 0.01 eV^2$, $D_{inel} = 0.07 eV^2$. The simulated electron mean free path in the multilayer system is about $1.48 \mu m$.

Apparently, from Fig. 4, electron-phonon scattering suppresses both NDR mechanisms and therefore, PVR values of these NDR peaks are reduced. However, the suppression of mechanism 2 is not as substantial as that of mechanism 1 as the quantum interference is more vulnerable to decoherence introduced by electron-phonon scattering. This might explain the absence of NDR peaks due to mechanism 1 in the experiments of [35]. However, in an experiment with sufficiently smaller devices, both mechanisms leading to the multiple NDR peaks can occur.

## 4. Size scaling analysis

System dimensions are the key ingredients in engineering the device performance. In this particular multilayer heterostructure, for instance, the device width determines the number of subbands in ANGR electrodes and the heterostructure length determines the length of interference region. Based on the two distinct mechanisms responsible for the multiple NDR peaks, it is intuitive that the device dimensions have significant and non-trivial influence on the NDR features rather than simply tuning the current magnitude by following quantum mechanical rules or Ohm's law. In order to comprehend such influences, the scaling analysis of the device dimensions, namely the lateral $(x, y)$ dimensions which defines the overlap area between hBN and graphene layers and the $z$-direction (number of hBN layers), is performed in this section. However, we do not consider the electron-phonon scattering effects during this analysis as they do not significantly alter the outcomes of the analysis.

## 4A. Tunneling barrier thickness ($N_z$)

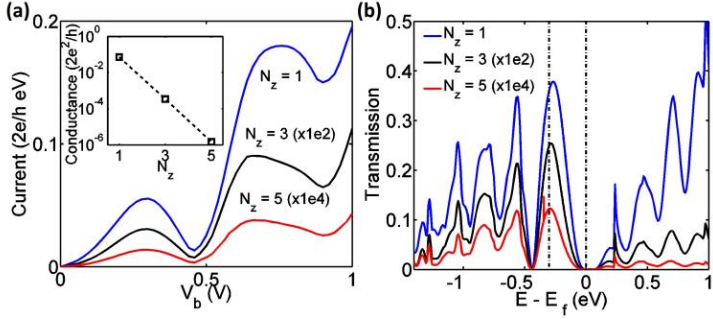

*Figure 5.* (a) Current-voltage curves for devices with different $N_z$, with fixed $N_x = 62$ and $N_y = 32$. Here the current value for cases when $N_z = 3$ and $N_z = 5$ are scaled by 1E2 and 1E4 respectively. The inset plots the low bias conductance of the three current-voltage curves. (b) Transmission relationship for devices with different $N_z$ and fixed $N_x = 62$ and $N_y = 32$ at $V_b = 0.3V$, corresponding to the first current peaks shown in (a). Again, the transmission coefficient value for cases when $N_z = 3$ and $N_z = 5$ are scaled by 1E2 and 1E4 respectively.

Representing the thickness of tunneling barrier, $N_z$ homogeneously modifies the current magnitude at different applied voltages, whereas has little effects on the peak positions and corresponding PVR. In Fig. 5(a), the hBN thickness ($N_z$) is varied from 1 layer (0.6nm) to 5 layers (2nm), while both $N_x$ and $N_y$ are fixed. The magnitudes of current are scaled by a multiplicative factor to present results on the same plot for different values of $N_z$. The transmission versus energy (Fig. 5(b)) shows that while the magnitude of transmission depends strongly on $N_z$, the locations of peaks depend weakly on $N_z$. Note that the dependence of current magnitude on $N_z$ lose its validity in the case when all the incident modes can tunnel through a thin barrier. This is because a thinner barrier only increases the tunneling probability of electrons without affecting the number of incident modes.

## 4B. Width of AGNR ($N_x$)

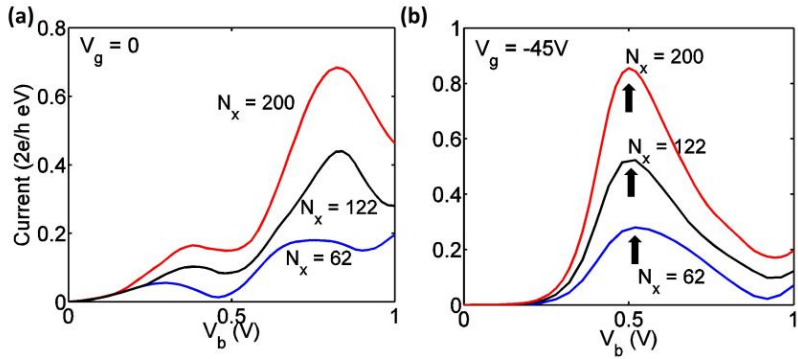

*Figure 6.* (a) Current versus drain voltage at $V_g = 0$ for devices with various $N_x$, with fixed $N_y = 32$ and $N_z = 1$. (b) Current versus drain voltage at $V_g = -45V$ for devices with various $N_x$, with $N_y = 32$ and $N_z = 1$. Black arrows mark the NDR peaks due to mechanism 2.

For the mechanism 1 (at $V_g = 0$), the density of subbands for the monolayer AGNR electrodes and the heterostructure region depends on the graphene nanoribbon width, *i.e.* the energy intervals between subbands for the structure with $N_x = 200$ are about three times smaller than that for the structure with $N_x = 62$. Therefore, a larger number of subbands contribute to current under lower biases, resulting in initial increase in current with $N_x$, as seen in Fig. 6(a). When the gate voltage is -45V, the NDR peaks induced by the mechanism 2 are observed near $V_b = 0.45V$ for different $N_x$ in Fig. 6(b). The heights of these peaks increase with device widths because the number of subbands carrying current between $\mu_B$ and $\mu_T$ grows with the width of the AGNR electrode (inset of Fig. 2(a)). We summarize the peak currents and PVR values for both mechanisms in Table I. Although the peak current is larger for the wider device, a rapidly decreasing PVR value can be observed. This is because of the stronger band-to-band tunneling between two AGNR contacts with a larger width, arising from the smaller subband spacing.

Table I exhibits a PVR value up to 13 for $N_x = 62$, which can be further increased to over 60 when $N_x$ shrinks to 14, showing the potential for the heterostructure to be utilized in both digital logic and memory. However, in reality to achieve the large PVR values will require a downscaling of $N_x$ and minimization of decoherence.

## 4C. Length of the heterostructure ($N_y$)

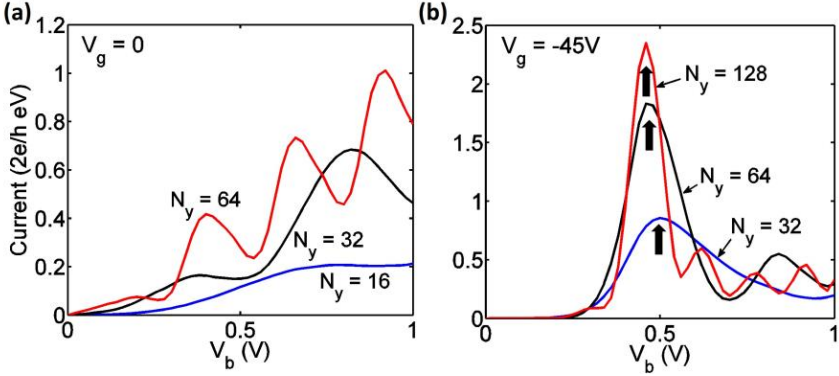

*Figure 7.* (a) Current versus drain voltage at $V_g = 0$ for devices with various $N_y$, with fixed $N_x = 200$ and $N_z = 1$. (b) Current versus drain voltage at $V_g = -45V$ for devices with various $N_y$, with $N_x = 200$ and $N_z = 1$. Black arrows mark the NDR peaks due to mechanism 2.

The length of the central multilayer region determines the number of incident carriers, and also characterizes the size of the Fabry-Pérot like interference cavity. For mechanism 1, when the heterostructure length $N_y$ changes from 16 (3.4nm) to 64 (13.6nm), the number of transmission peaks increase as shown in Fig. 7(a). The NDR peaks appear at $V_b = 0.38V$, 0.8V for $N_y = 32$, which shifts to lower $V_b$ i.e. at 0.2V, 0.4V respectively, when $N_y = 64$. This is because the resonant transmission appears at various energies, which vary inversely proportionally with the length of the interference cavity $N_y$. Experiments where the overlap between two graphene nanoribbons are altered should be able to reveal the differences in oscillations of I-V characteristics as a function of $N_y$. We note that experiments with changing overlap have been performed in carbon nanotubes before [44,45] and future experiments in BN-graphene heterostructures should be useful in studying these features.

With an ultrathin tunneling barrier ($N_z = 1$), electrons have high tunneling probabilities and thus the current is mainly limited by the number of modes incident within the energy window. Graphene layer has low DOS near Dirac point, yielding the saturation of peak current at large $N_y$. It is also observed that for larger $N_z$ ($N_z \geq 3$, results not shown), the peak current increases with $N_y$ rapidly without saturation. This is consistent with our previous discussion since a thicker hBN tunneling barrier greatly suppresses the overall tunneling transport probability and therefore the peak current magnitude is a strong function of the number of carriers in graphene.

*Table I: peak current and PVR values as a function of $N_x$ for both mechanisms. (I-V curves from Fig. 6)*

| $N_x$ ($V_g = 0$) | 62 | 122 | 200 |
|---|---|---|---|
| Peak Current ($2e^2/h$) | 0.05 | 0.10 | 0.16 |
| PVR | 4.2 | 1.19 | 1.06 |
| | | | |
| $N_x$ ($V_g = -45V$) | 62 | 122 | 200 |
| Peak Current ($2e^2/h$) | 0.28 | 0.53 | 0.85 |
| PVR | 13 | 5.9 | 5.0 |

## 5. Conclusions

We have systematically investigated the charge transport properties of a three-terminal graphene-hBN-graphene multilayer heterostructure device as a function of device dimensions, so as to further understand the underlying mechanisms for negative differential resistance. The prototypical graphene-hBN-graphene multilayer heterostructure has two distinct mechanisms that can introduce NDR behavior. The first mechanism involves a Fabry-Pérot like resonant feature due to interference in the multilayer heterostructure region, which can produce multiple current peaks. In the presence of an external gate, resonant tunneling can also occur when the electronic spectrum (Dirac points) of the top and bottom graphene electrodes align, which leads to a second mechanism for resonant tunneling. Both mechanisms respond to gate voltage distinctly. Gate voltage only controls the locations of NDR peaks from mechanism 1 while can tune both PVR and locations of NDR peaks due to mechanism 2. More interestingly, in the presence of electron-phonon scattering decoherence, mechanism 1 is more intensively suppressed compared to mechanism 2, which explains the disappearance of mechanism 1 in the experiments.

Size scaling analysis provides insight into the device physics that determines the number of NDR peaks, the variation of peak current and PVR value with change in device dimensions. (1) The hBN thickness exponentially controls the magnitude of current without significantly affecting the NDR features. (2) For devices with larger widths ($N_x$), the multiple current peaks preserve but with decreasing PVR values for both mechanisms. (3) For mechanism 1, the bias voltages at which multiple current peaks, the number of peaks increase with length. In contrast to this, location of single peak originated from mechanism 2 is independent of the length. We believe that the negative differential resistance's sensitivity to the system dimensions will provide additional insights for future theoretical and experimental investigations.